\documentstyle[preprint,tighten,aps]{revtex}

\begin{document}

\preprint{\vbox{\noindent
%Submitted to TBA
%PRELIMINARY DRAFT
          \null\hfill  INFNFE-09-97}}

\title{A MIXED SOLAR CORE, SOLAR NEUTRINOS AND HELIOSEISMOLOGY}

\author{
         S.~Degl'Innocenti$^{1,2}$ 
         and B.~Ricci$^{2}$
       }
\address{
$^{1}$Dipartimento di Fisica dell'Universit\`a di Pisa, Piazza Torricelli 2, I-56100 Pisa, Italy\\
$^{2}$Istituto Nazionale di Fisica Nucleare, Sezione di Ferrara,
      via Paradiso 12, I-44100 Ferrara, Italy\\
}
\date{October 1997}
\maketitle                 % Produces the title.

\begin{abstract}

We consider a wide class of  solar models with mixed core.
Most of these models can be excluded as the
predicted sound speed profile is in sharp disagreement 
with helioseismic constraints. 
All the remaining models  predict $^8$B and/or $^7$Be neutrino fluxes
at least as large as those of SSMs.
In conclusion, helioseismology shows that a mixed solar core 
cannot account for the neutrino deficit implied by the current solar
neutrino experiments.
\end{abstract}

\section{Introduction} 

The  hypothesis of a mixed solar core was advanced in the seventies as
a desperate attempt to solve the solar neutrino puzzle of
that time, i.e. the low signal reported by the Chlorine experiment in
comparison with the Standard Solar Model (SSM) prediction
\cite{EC68,SS68,BBU68}.

Roughly speaking (see later for more precise statements) in mixed
solar models the hydrogen content of the innermost region is
enriched, so that nuclear fusion gets easier and the observed solar
luminosity can be obtained at smaller central temperature.
Correspondingly the $^8$B and $^7$Be neutrino
fluxes should be strongly decreased.

It was clear since the beginnings that mixing was a hardly tenable hypothesis
\cite{SS71}. The Sun is a typical population I star and advanced
evolutionary phases of solar like stars are observed in galactic open
clusters. In the presence of an extended mixed
core, the evolution of stars off the main sequence is altered in
disagreement with the observed color magnitudine diagram of open
clusters.
 However, the proposal of a mixed solar core
could not be completely discarded in the absence of 
direct observational constraints on the internal solar
structure.

In this respect, helioseismology provides a powerful
tool \cite{UR83}. Recently we have shown \cite{DZFR97}
that the sound speed near the solar center
(R/R$_{\odot} \approx 0.1 -0.2$) can be determined 
with an accuracy of few percent or even better.
The predictions of mixed core models (MCMs) can thus 
be confronted with observational data.
The results of a  systematical analysis are
presented in this note.

We remind that quantitative estimates of the neutrino fluxes predicted
by MCMs were presented in \cite{BBU68,SBP90}.
Interest on this matter was revived recently  by Cummings and Haxton 
\cite{CH97}:  by
considering an artificial $^3$He mixing with different upward and downward flow
velocities, they found  that the discrepancy between solar neutrino
observations and predictions can  be reduced.  Bahcall et al. \cite{BPBCD97}
have shown however that this model yields a sound speed profile
drastically different from that inferred by helioseismology. In a
similar context, Richard and Vauclair \cite{RV97} 
considered  local mixing induced by an anomalously large diffusion
near the edge of the nuclear burning core, showing that the resulting
models are  incompatible with
helioseismic data.

With respect to the previous literature, the main features of   this paper are
the following ones:\\

\noindent
i) we take into account quantitatively the accuracy of the sound speed
inferred from helioseismic data: in addition to the measurement
errors, we consider the systematic uncertainties due  to the
choice of the starting solar model (used in the linearized
inversion technique) and  to the free parameters of the inversion
algorithm.\\

\noindent
ii) We systematically analyse sound speed profiles from
fast and slow mixing models, for a wide choice of
mixing region. We discuss both continuous and episodic 
mixing processes.\\

\noindent
iii) As a result, we show that all  MCMs which  cannot be
ruled out yield $^8$B and $^7$Be neutrino fluxes at least as large 
as in SSMs.

\section{ The sound speed profiles of mixed solar models}

We have studied the evolution of stellar structures 
by using the latest version of FRANEC \cite{CDR97}.
Artificially, the internal composition is mixed,
up to a radius $R_{mix}$. We have adjusted the original 
chemical composition and the mixing length so as to 
produce at the solar age ($t_\odot =4.57$ Gyr \cite{BP95})
the observed solar
properties: luminosity, radius and photospheric
Z/X. In this way we have produced (non-standard) solar models.

Firstly, we discuss in some detail the case of continuous
mixing, i.e.  the solar core is assumed to be  mixed 
continuously from ZAMS to present.

We distinguish two regimes, by comparing the 
circulation time $t_{circ}$, i.e. the time needed for material 
circulation in the core, with the time scale of $^3$He equilibrium 
abundance, $t_3 \approx 10^7$ yr.
Following the
classification of Ref. \cite{SS68}, we call ``slow" (``fast")  mixing 
that characterized by $t_{circ} >> t_3$ ($t_{circ}  << t_3 $).

In the slow mixing case
 the $^3$He abundance is determined locally by its
equilibrium value, while the abundances of H and  $^4$He are 
kept  uniform  in the mixed region.
For the fast mixing, H, $^3$He and $^4$He  are all kept uniform.

We anyhow assume that $t_{circ}$ is much larger than the 
radiative transport time $t_{r} \approx 10^6 $ yr 
(in the language of Ref. \cite{SS68}, we  neglect 
``superfast'' mixing).

Concerning the extension of the mixing regions, calculations
have been carried out in all cases for
 $R_{mix}$=0.05, 0.1, 0.15, and 0.2 $R_\odot$.

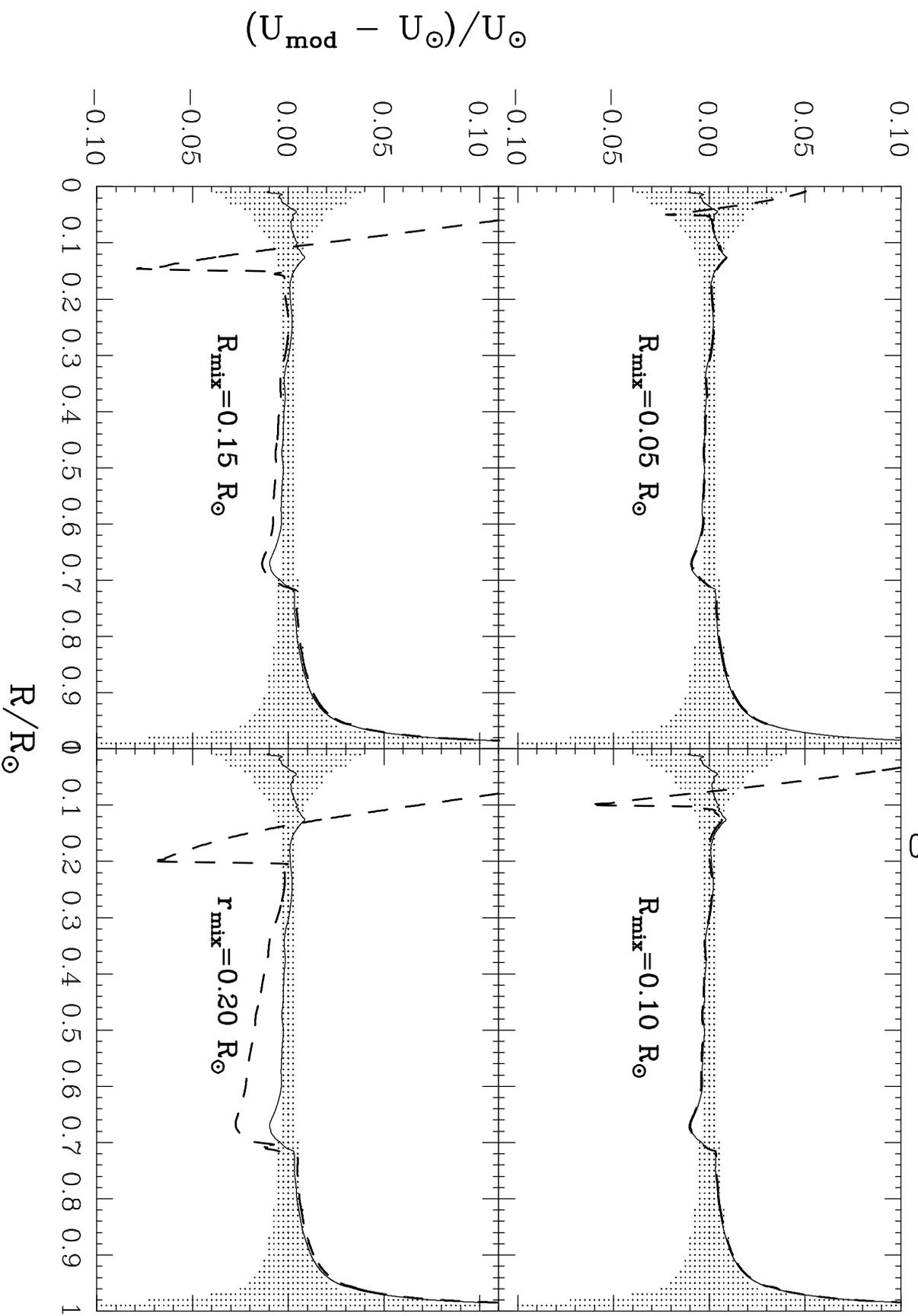
\begin{figure}
\caption [a]{ 
For the indicated values of $R_{mix}$, we present the relative difference
between the isothermal sound speed as predicted by solar model with fast 
continuous mixing, U$_{mod}$, and the helioseismic determination,
U$_{\odot}$.
The dotted area corresponds to the uncertainty
on U$_{\odot}$.
}
\label{figumixf}
\end{figure}

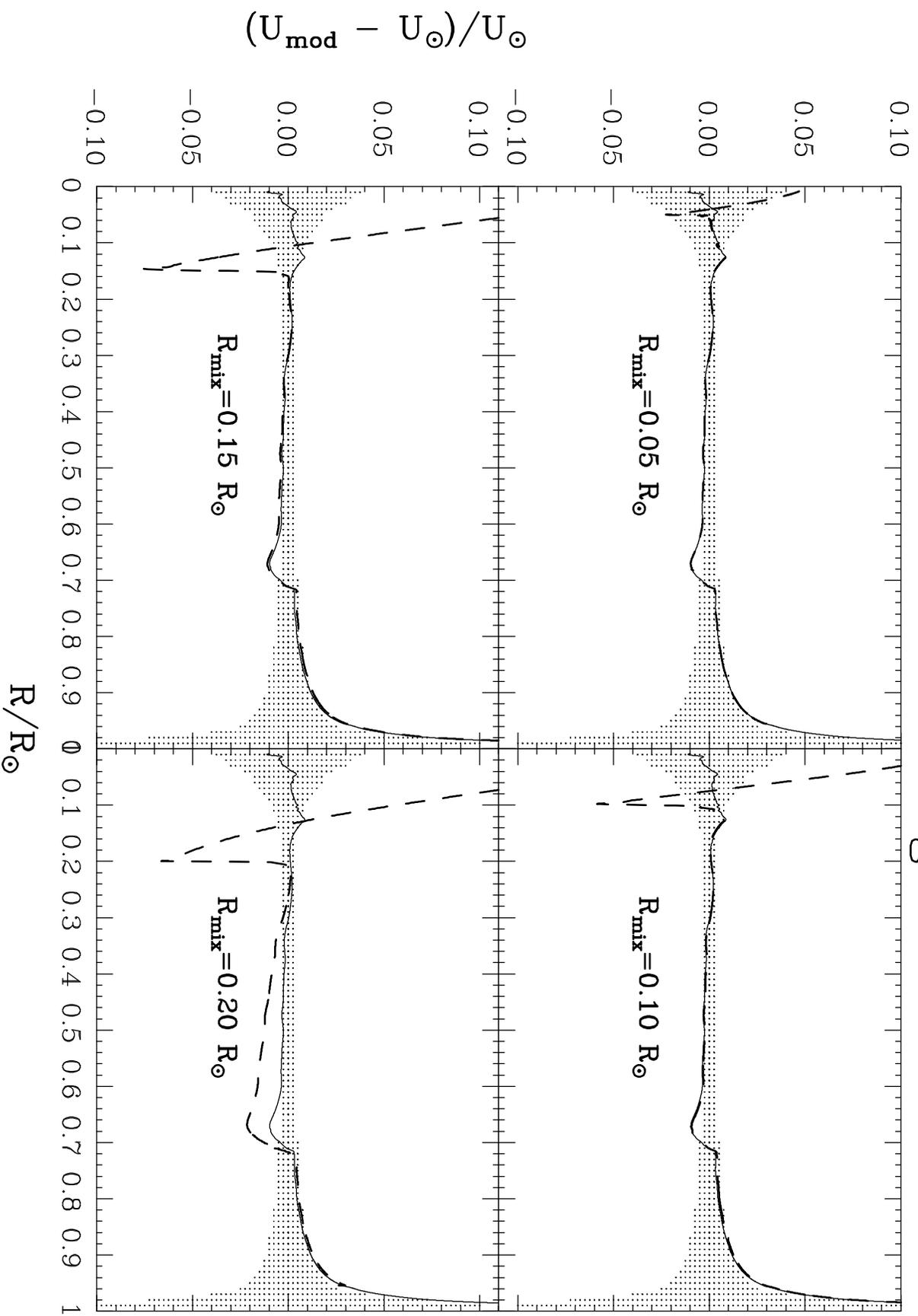
\begin{figure}
\caption [a1] {The same as in Fig. \ref{figumixf}, but for slow continuous
 mixing. $\quad \quad \quad \quad \quad \quad \quad \quad \quad \quad \quad \quad $}
\label{figumixs}
\end{figure}

In Figs. \ref{figumixf} and \ref{figumixs} we show (dashed line) 
 the relative difference between the isothermal sound speed
squared, U=P/$\rho$, as predicted by  the mixed models and the 
helioseismic  determination, U$_\odot$. The same quantity for
our SSM is also  shown (solid line).

The dotted area corresponds to the conservative uncertainty on U$_\odot$, 
as in  ref. \cite{DZFR97}. 
We remark that the accuracy of the 
helioseismic determination, although degrading at very small radii, is
still better than 1\% at $R/R_\odot\simeq 0.1$. 
Most of this (conservatively estimated) error 
 \footnote{We remind that our 
estimate of uncertaintes is very conservative because
 we add linearly the contibution 
of observational errors and of the systematical errors due to the 
inversion technique}
arises from the uncertainties
of the inversion method; the observational errors on the measured 
frequencies contribute just an uncertainty  $\Delta$U/U$\simeq$ 0.1\%.

The predictions of acceptable solar models have to lie within the 
dotted area in Figs. \ref{figumixf} and \ref{figumixs}, and 
actually our SSM (solid line) is generally in agreement with 
the helioseismic constraint. 

On the other hand 
one sees a strong deviation of MCMs (with respect to U$_\odot$ and 
U$_{SSM}$) in the mixed zone; this is a consequence of 
the change in ``mean molecular weight'', $\mu$,  induced by mixing.
In the approximation of perfect gas  
(accurate to the level of few per thousand  in the solar core)
one has U$\propto T/ \mu$. Due to mixing, the innermost region 
is enriched with hydrogen, so that $\mu$ decreases 
(we observe that change of $\mu$ can be as high as 40\%,
whereas temperature change is at most  a few per cent) and
U increases. The opposite occurs near the edge of the mixed
region.

As the mixing area increases the sound speed profile of the
MCMs deviates more and more from the SSM prediction and
it becomes  in conflict with helioseismic constraint if $R_{mix} \ge 0.1 R_\odot$.

We remark that the sound speed profile is altered also
outside the mixing region (see Figs. \ref{figumixf} and 
\ref{figumixs} for $R_{mix}=0.2 R_\odot$),
however this effect is weaker and provides a weaker constraint on $R_{mix}$.

Episodic mixing is another possibility which is worth
studying, see also Ref. \cite{SBP90}. One assumes that once
in the solar history a violent redistribution of material
occurs, in a time shorter than $t_3$.
Episodic mixing can alter significantly the subsequent stellar evolution, as 
its effects will be washed out only on a nuclear 
burning time scale, a few Gyr.

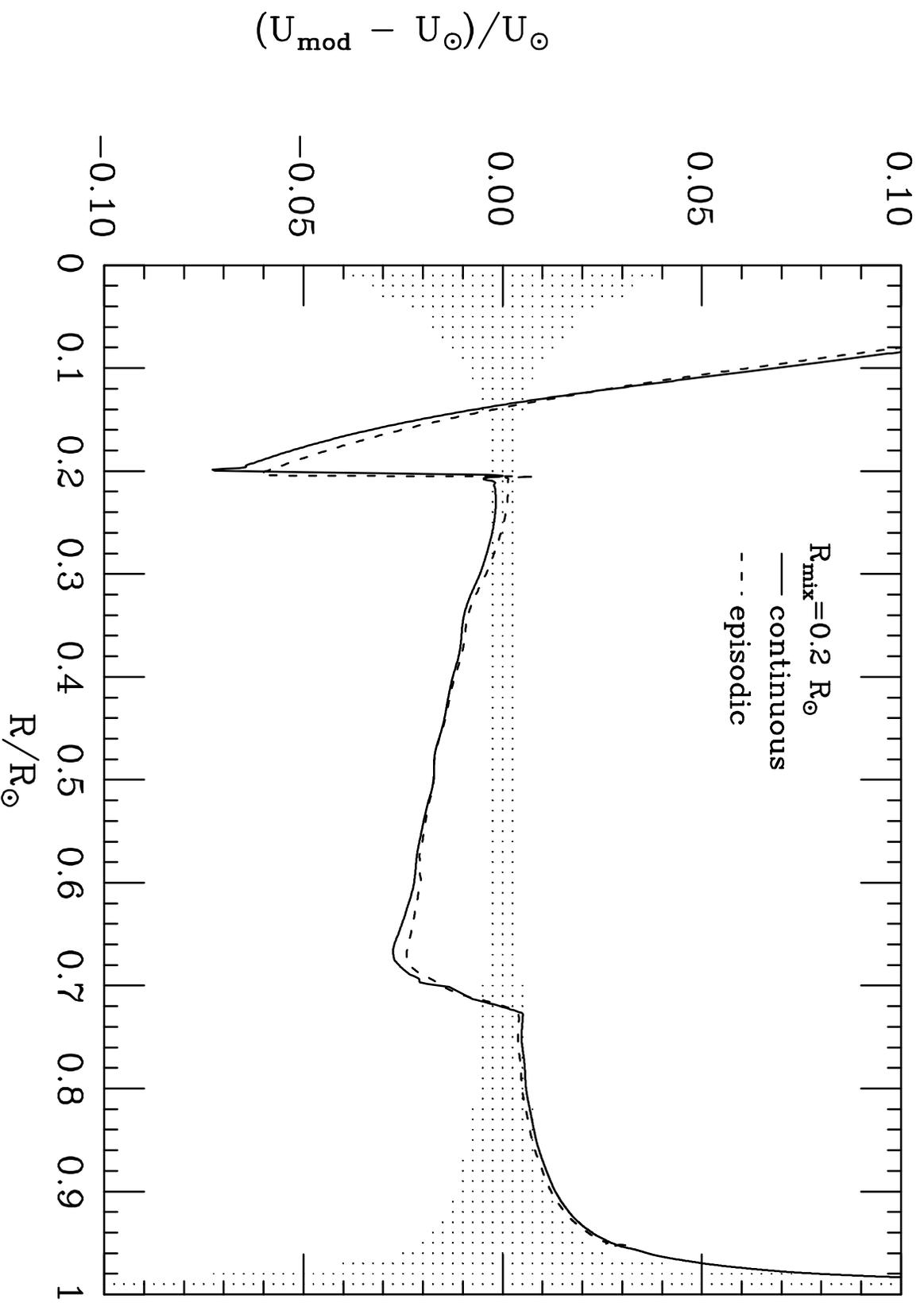
\begin{figure}
\caption [aa] {The relative differences 
(U$_{mod}-$U$_\odot)/$U$_\odot$ for  solar models 
with fast continuous and  episodic mixing (at $t=4.54$ Gyr), both for 
a mixed region extending up to R$_{mix}=0.2$ R$_\odot$. 
The dotted area corresponds to the uncertainty 
on U$_\odot$.
}
\label{figepi}
\end{figure}

If episodic mixing occured early in the solar hystory, its effects
are cancelled both on stellar structure and on neutrino 
fluxes, see Ref. \cite{SBP90}.
On the other hand, a recent mixing should yield a stellar structure
similar to that of continuous fast mixing.
This expectation is confirmed by numerical calculations. In
 Fig. \ref{figepi} we present the results of MCMs for $R_{mix}=0.2 R_\odot$,
both for continuous fast mixing and for an episodic mixing occuring
at $t=4.54$ Gyr after ZAMS.

In conclusion, for the comparison with helioseismology the results of 
continuous fast mixing can be considered as representative of recent
episodic mixing too.

\section{ Neutrino fluxes}

 As the neutrino fluxes are strongly dependent on 
the central solar temperature $T_c$ (see e.g. Refs.\cite{Report,libro}), we
first study  the values of $T_c$ in  the continuous
MCMs we have considered.

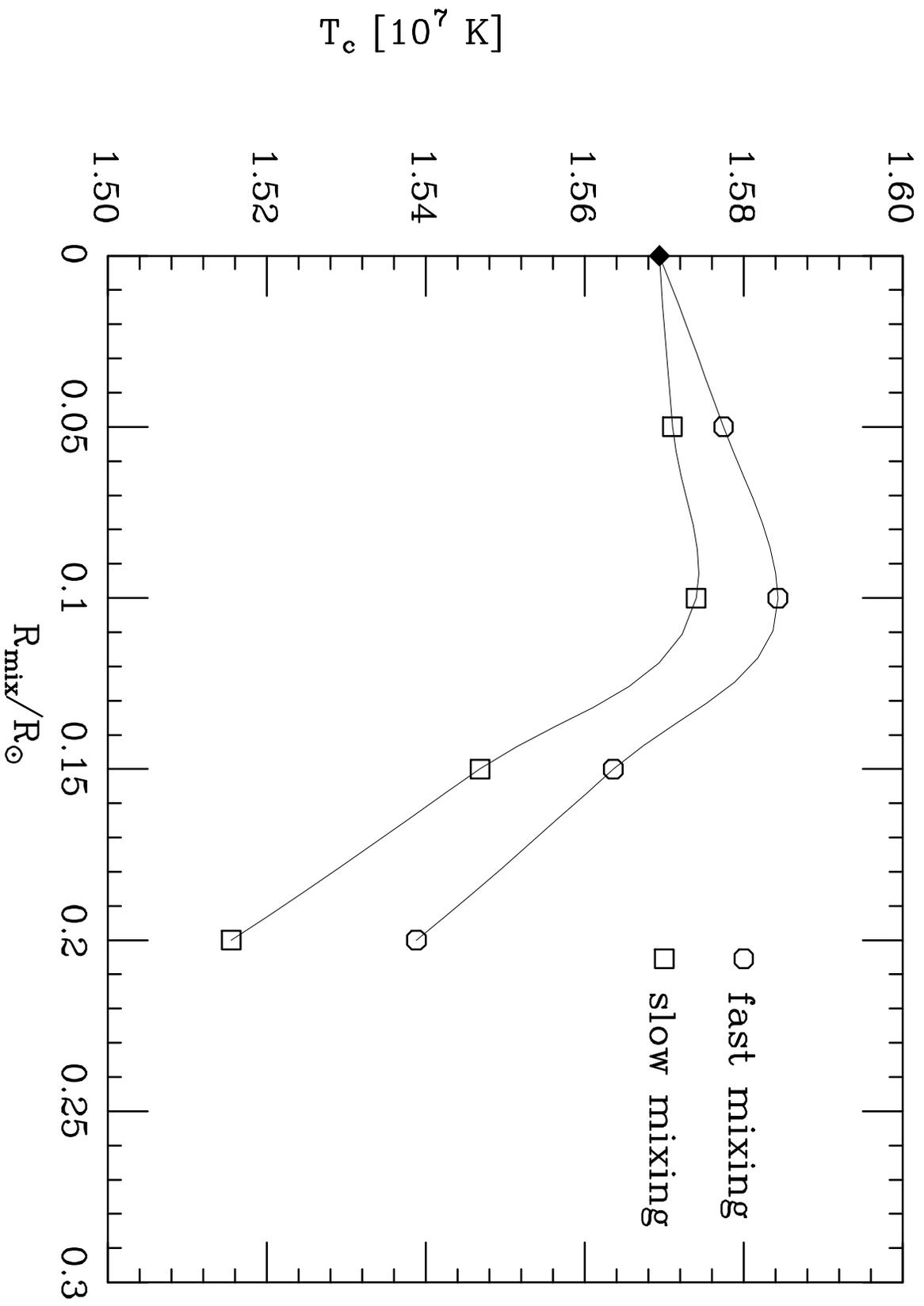
\begin{figure}
\caption [a2] {The central temperature $T_c$ 
of models with fast and slow continuous
mixing as a function of the mixing radius R$_{mix}$. }
\label{figTc}
\end{figure}

The behaviour of $T_c$ as a function of the mixing radius (see Fig. 
\ref{figTc}) has been already exhaustively
discussed in Ref.  \cite{SS68} for the fast MCMs.
 We find that also the slowly MCMs
start increasing the central temperature, an
occurrence which deserve some comments (being for slow mixing $^3$He
 at local equilibrium values, the mechanism discussed in  Ref. \cite{SS68}
cannot be at work).

Numerical experiments have clearly demonstrated
that the case of slow mixing is governed by two conflicting
mechanisms:\\
ii) the occurrence of a mixed region shifts
the energy production at the center of the structure and consequently a
smaller fraction of stellar matter is producing energy. To produce
the same amount of energy the central temperature must increase. Note
that this mechanism is reinforced by the fact that the energy is
produced at smaller radii, so that the radiative flux (for a given energy
output) is increased, increasing in turn the radiative gradient
and as a consequence
the temperature around the center quickly drops below the limit for
efficient nuclear reactions. \\
ii) On the other hand, as in  mixed models the core is enriched 
with hydrogen, nuclear fusion gets easier and 
this acts in decreasing the temperature. 

From numerical experiments it is clearly shown 
that for a limited amount
of mixing ($R_{mix} \leq 0.1 R_\odot$) the first mechanism dominates and the central
temperature increases. For larger mixing the increase of central H overcomes 
the first mechanism (i.e. the shift of the burning toward the center)
and the central temperature starts decreasing.

%%%%%

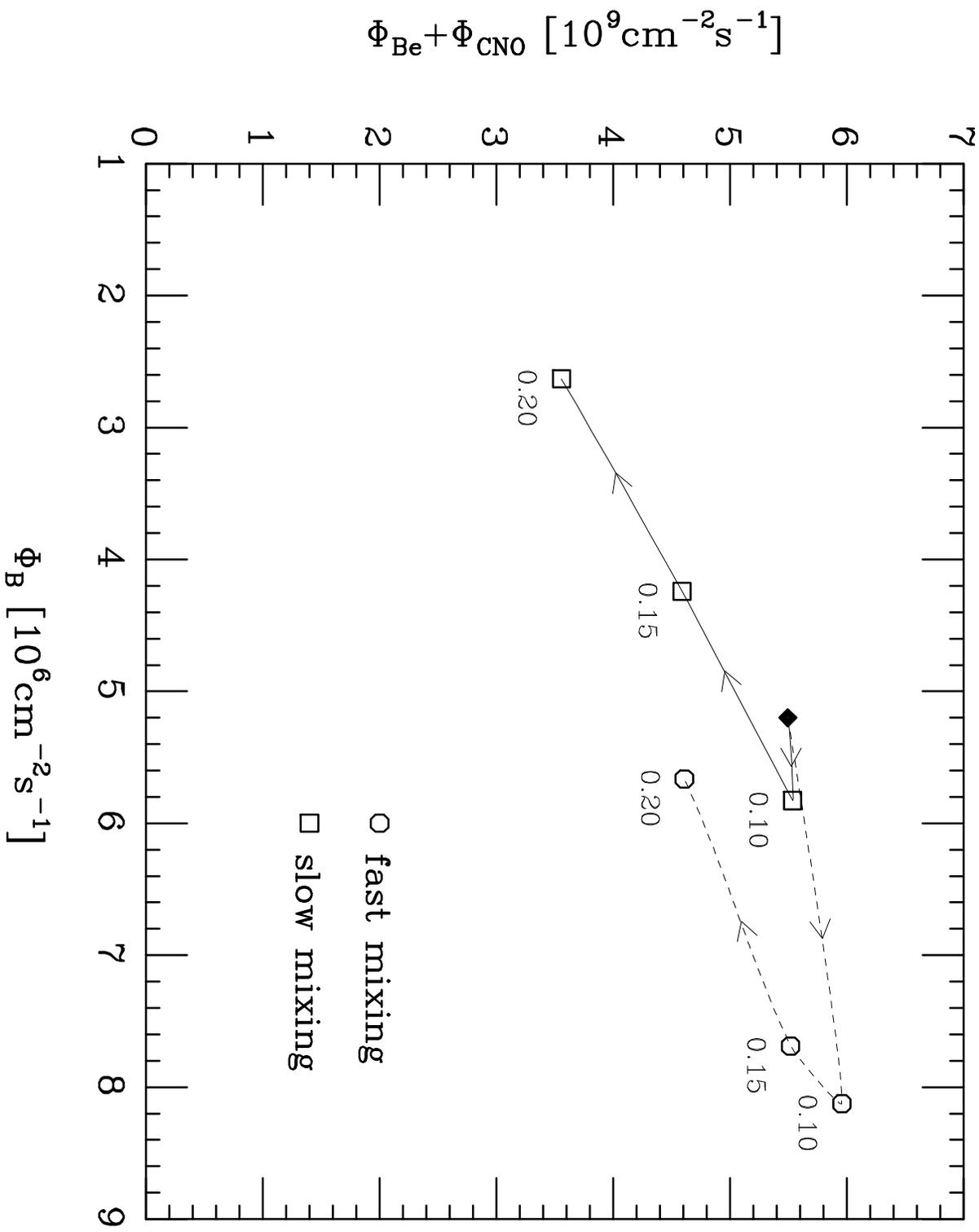
\begin{figure}
\caption [b] {
The predictions of the intermediate 
($^7$Be+CNO) and high energy ($^8$B) neutrino fluxes in  solar models 
with continuous mixing, for the indicated values of $R_{mix}$.
The prediction of our SSM (full diamond) is also shown.
}
\label{figbebo}
\end{figure}

The basic features of neutrino fluxes 
predicted by MCMs are shown in Fig. \ref{figbebo}.
They reflect the temperature behaviour mentioned above:
if the mixed region is small enough both
 the intermediate energy ($^7$Be+CNO) and high
energy ($^8$B) component are larger than the SSM prediction.
Reduction of these components is obtained 
only for rather extended mixing, which, as shown in the previous section,
are inconsistent with helioseismic constraint.

In conclusion helioseismology shows that a mixed solar core cannot
account for the deficit of intermediate and/or high energy neutrino component, 
as implied  by the current solar neutrino experiments 
(see e.g. Ref. \cite{Report}).

\acknowledgments
We are extremely grateful to V. Berezinsky,  V. Castellani, W. A. Dziembowski
and G. Fiorentini for their useful suggestions and  comments.

\end{document}